\documentclass[10pt,a4paper]{article}
\usepackage{amssymb,latexsym,amscd,amsmath,epsfig}
\def\be{\begin{equation}}
\def\ee{\end{equation}}
\def\ba{\begin{array}}
\def\ea{\end{array}}
\def\bea{\begin{eqnarray}}
\def\eea{\end{eqnarray}}
\def\bi{\begin{itemize}}
\def\ei{\end{itemize}}

\begin{document}
\textwidth=135mm
 \textheight=200mm

\begin{center}
{\bfseries A New Study of the Transition to Uniform Nuclear Matter
in Neutron Stars and Supernovae \footnote{{Participant
Contribution at the ``Dense Matter in Heavy Ion Collisions and
Astrophysics" Summer School, JINR, Dubna, Aug. 21 - Sept. 1,
2006.}}}

\vskip 5mm

W. G. Newton$^{\dag}$

\vskip 5mm

{\small {\it $^{\dag}$ University of Oxford, Condensed Matter
Dept., Clarendon Laboratory, Parks Rd., Oxford, OX1 3PU, U.K.}} \\

\end{center}

\vskip 5mm

\centerline{\bf Abstract} A comprehensive microscopic study of the
properties of bulk matter at densities just below nuclear
saturation $\rho_s = 2.5 \sim 10^{14}$ g cm$^{-3}$, zero and
finite temperature and high neutron fraction, is outlined, and
preliminary results presented. Such matter is expected to exist in
the inner crust of neutron stars and during the core collapse of
massive stars with $M \gtrsim 8M_{\odot}$

\vskip 10mm

\section{\label{sec:intro}Introduction}

Understanding the phase transition from inhomogeneous to uniform
nuclear matter is important in the study of a number of
astrophysical phenomena. In the region of density $0.05\rho_s
\lesssim \rho \lesssim 0.16\rho_s$, temperature $0 < T \lesssim
10$MeV and proton fraction $0.01 \lesssim y_p \lesssim 0.3$, which
is expected at the bottom of neutron star crusts and during core
collapse supernovae, heavy nuclei immersed in a fluid of neutrons
are expected to become distorted into a series of exotic
structures known as `nuclear pasta' \cite{r1} in order to minimize
the sum of their surface tension energy and Coulomb repulsion with
adjacent nuclei. These structures may extend over distances many
times `normal' nuclear radii $\approx$ 10fm.

Such matter could have a significant impact on the dynamics of
core collapse supernovae. It exists in the outer regions of the
collapsed core where neutrinos are expected to be trapped. The
nuclear pasta phases may affect the neutrino opacity of the matter
as excitation of collective modes in the pasta offers another
channel for the transfer of energy from neutrino flux to the
nuclear medium \cite{r2}.

In the inner crust of neutron stars the pasta phases are expected
to be a form of `soft' condensed matter, that is, in a state
between liquid and solid \cite{r3}. In addition, the external
neutron gas is expected to be in a superfluid state, and as such
its bulk flow will be quantized into vortices, whose cores are
comparable to the extent of the pasta shapes in size \cite{r4}.

Our theoretical studies are required for the explanation of
observations. Thermal emission has been observed from seven young
neutron stars \cite{r5}; to constrain models of the cooling of the
neutron stars' cores, we must know the relevant mechanisms of heat
transport in the crust and their efficiency. Quasi-periodic
oscillations have recently been observed in the tails of flares
from soft gamma-ray repeaters \cite{r6} and have been interpreted
as being produced by seismic waves in the crust as it relaxes
after the significant readjustment that triggered the flare.
Glitches in the spin down rates of pulsars tell us something about
the coupling of the crust to the core \cite{r7}. Later in a
neutron stars' life, accretion may lead to the formation of
`mountains' on the neutron star surface, leading to a quadrupole
moment and gravitational wave emission: whether the gravitational
waves can be detected on Earth depends on how big a mountain the
crust can support \cite{r8}. Accretion can also reheat the crust,
melting it in layers \cite{r9} and will also compress the
innermost layers of the crust.

\section{\label{sec:ch1} Motivation for a New Study of the Pasta Phases}

In order to study the structure of the inner crust in an unbiased
way, one must perform the calculations self-consistently in three
dimensions, making no assumptions about the shape of the nuclear
distribution or lattice type (i.e. free of the Wigner-Seitz
approximation) and no distinction between the nuclear clusters and
external neutron gas. Because of the computationally intense
nature of such calculations, they have only recently been
attempted, by two methods in particular; the semi-classical
Quantum Molecular Dynamics (QMD) \cite{r11} and the fully quantum
mechanical Hartree-Fock (HF) method \cite{r12}. QMD allows for a
large volume (of order 100fm) to be simulated, and thus longer
range (lower energy) effects explored. HF is more intensive
computationally, so the simulation volumes are smaller (of order
20-40fm), but effects arising from the shell structure of the
nuclear clusters and the unbound nucleons which scatter off them
are automatically included. Both the above studies also neglected
the band structure of the unbound nucleons, for which one requires
general Bloch boundary conditions

We would like to extend the Hartree-Fock study above to a far more
comprehensive range of density and proton fraction space, as well
as extending it to finite temperature thus exploring the
properties of matter in core collapse supernovae. In doing this we
would survey the effects of shell structure on the energy and
ordering of the shapes, paying particular attention to separating
physical effects from numerical artifacts, and implementing the
general Bloch boundary conditions so we can examine the band
structure of the neutron fluid.

\section{3d Hartree-Fock Simulation of Bulk, Inhomogeneous Nuclear Matter}

The Hartree-Fock approximation assumes a definite set of
quasi-particle states with energies $\epsilon_1,...,\epsilon_A$
that are occupied with a probability given by the Fermi-Dirac
distribution

\be \label{e12} n_{i,q} = {1 \over e^{(\epsilon_{i,q} - \mu_q)/k_B
T}  + 1} \ee

where $q = p,n$ labels the isospin states, $i$ the single particle
states and $\mu$ is the chemical potential. Physically we are
making the assumption that the nucleons in a nucleus or a nuclear
configuration move independently of each other in an average
potential created by all the other nucleons. The variational
principle $ \label{e4} \delta E[\Phi] = \delta \langle \Phi|
\hat{H} | \Phi \rangle = 0 $ is used to obtain our approximation
to the ground state given a Hamiltonian containing a two-body
nuclear interaction. We use the Skyrme interaction, a zero-range
effective nuclear force particularly suited to Hartree-Fock
calculations. Carrying out the variational procedure with respect
to the single particle wavefunctions of a single Slater
determinant, the two body Skyrme potential becomes a one body
density-dependent potential and the A-body Schro\"dinger equation
for the A-body wavefunction becomes A single body Schro\"dinger
equations for the single particle wavefunctions:

\be \label{e5} h_{HF} \phi_{i,q} = \bigg[ - \vec{\nabla} {\hbar^2
\over 2 m_q^*} \vec{\nabla} + u_q(r) + w_q(r) {(\vec{\nabla}
\times \sigma) \over i} \bigg] \phi_{i,q} = \epsilon_{i,q}
\phi_{i,q} \ee

\begin{figure} [!t]
\centerline{\psfig{file=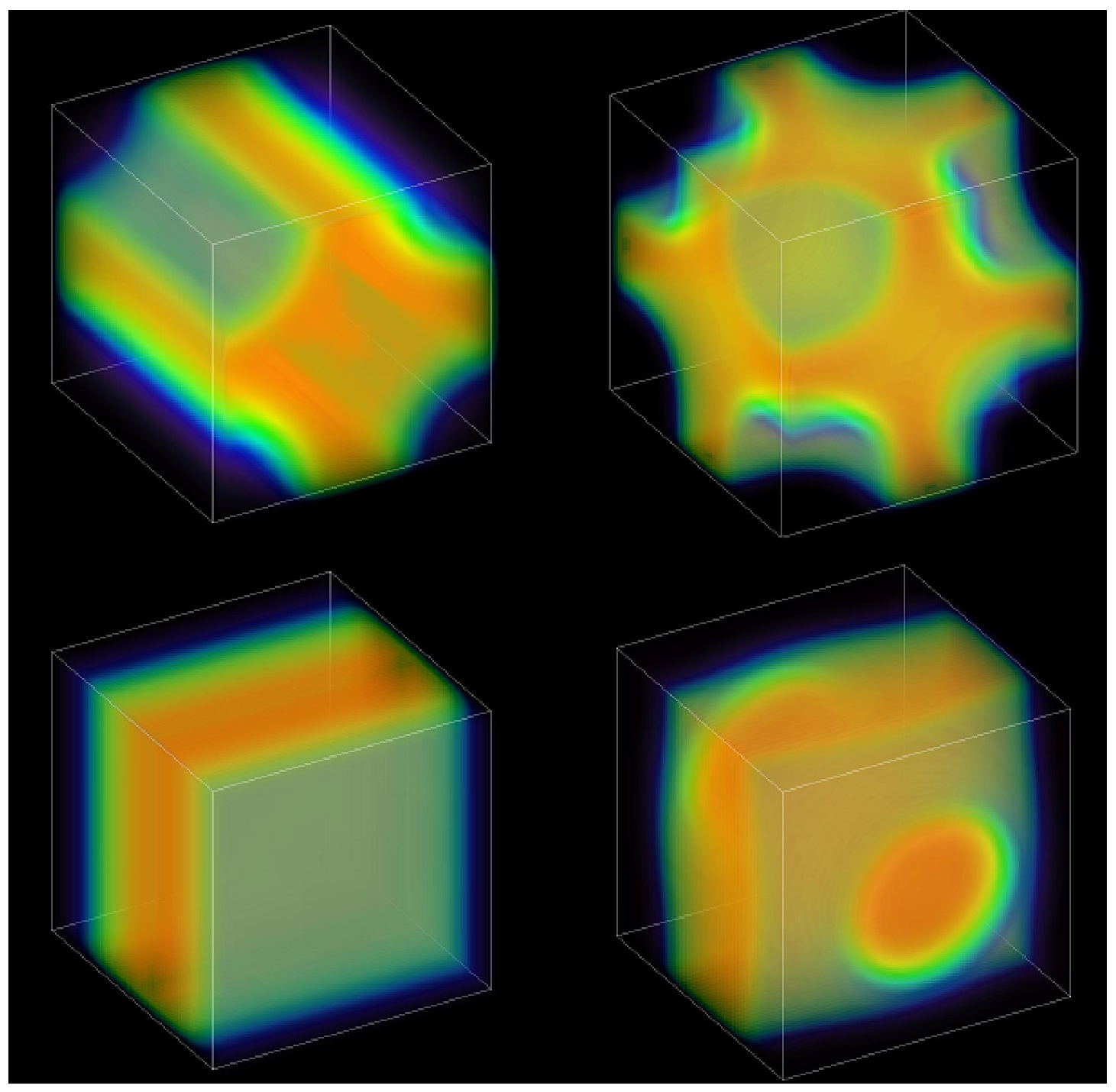, width=5.5cm}
\psfig{file=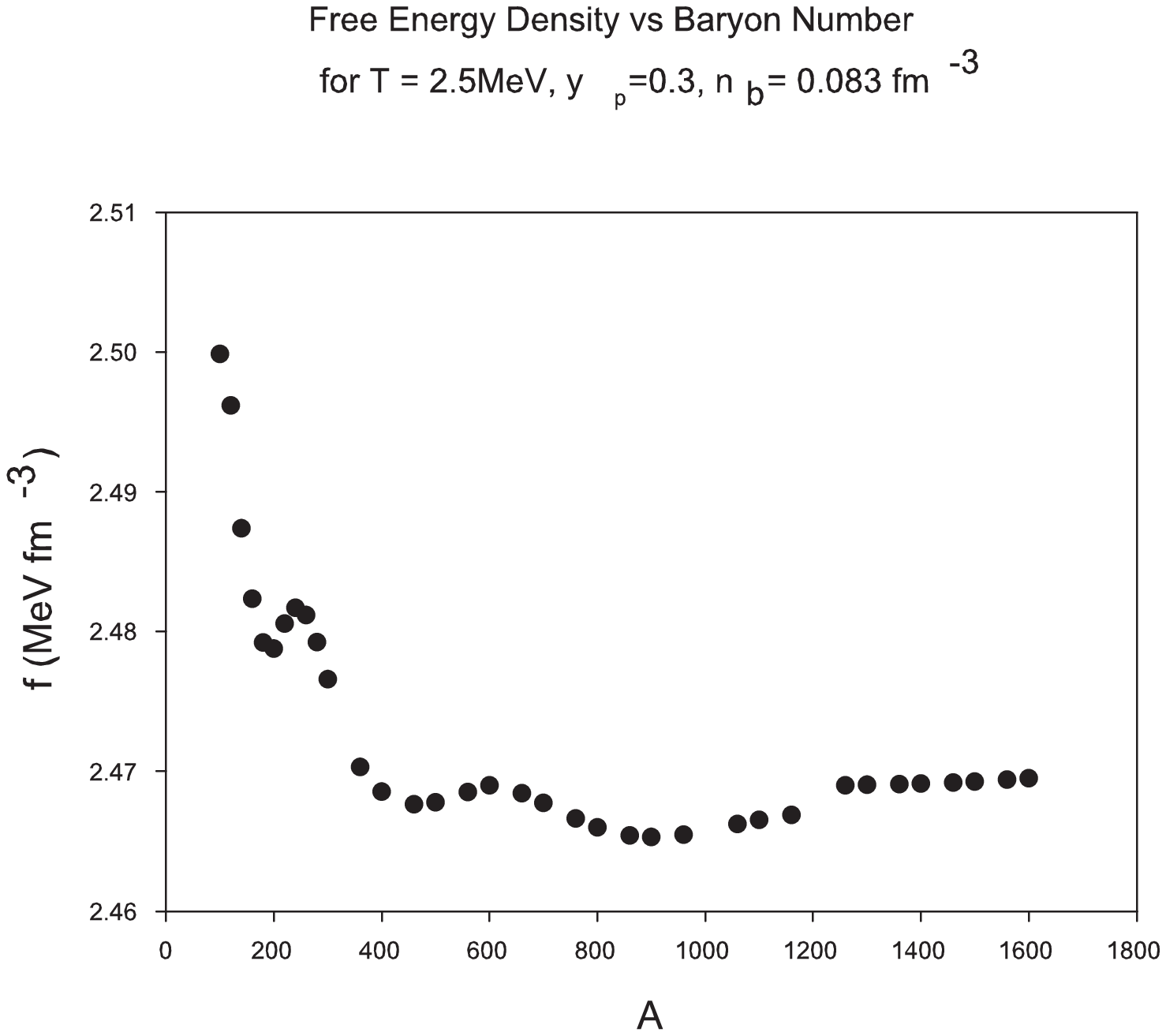, width=8cm}} \caption{Right: The total free
energy density, at temperature, density and proton fraction as
given, as a function of number of nucleons in the cell (or,
equivalently, the size of the cell). Left: some nuclear
configurations obtained: top left - A=460, top right - A=2200,
bottom left - A=280, bottom right - A=1300.} \label{f1}
\end{figure}

Here, $w_q$ is the spin-orbit potential (which we currently set to
zero), $u_q$ is the single particle potential, and $m_q^*$ is the
effective mass. See \cite{r10} for the form of the Skyrme
interaction and derived one body potentials.

We also impose a constraint that the neutron quadrupole moment of
the nuclear configuration be given by an input value $\langle Q
\rangle_p$. If we didn't do this, we would have no control over
which of the local minima in deformation space the simulation
would fall into, destroying self-consistency.

We make the assumption that at a given temperature and density the
matter is arranged locally in a periodic structure with an
indentifiable unit cell. We then take our computational volume to
be that unit cell. We impose Bloch boundary conditions:

\be \label{bloch} \phi_{i,q}(\vec{r} + \vec{T}) = e^{i\vec{K}
\cdot \vec{T}} \phi_{i,q}(\vec{r}) \ee

where $\vec{T}$ is the translation from the position $\vec{r}$ to
the equivalent positions in the adjacent cells, and $\vec{K}$ is
the Bloch momentum covector. We must perform one simulation for
each value of $\vec{K}$ within the first Brillouin zone.

Each unit cell will contain a certain number of neutrons N and
protons Z, making a total baryon number of A = N + Z. We have
freedom to increase the cell volume V and number of nucleons A and
still describe the same density.

Thus at a constant density, temperature and proton fraction we
must scan across different cell sizes, neutron quadrupole moments
and Bloch covectors. In order to reduce the numerical work, we
will actually scan across cell size and quadrupole moment with
simple periodic boundary conditions ($\vec{K} = 0)$, select the
configuration that gives the minimum energy, and then perform the
calculation for that configuration with $\vec{K} \neq 0$.

We are in the process of performing the calculations outlined
above on the Cray XT3 super computer Jaguar at Oak Ridge National
Lab. A sample of preliminary results is given in
figure~(\ref{f1}).

\end{document}